# Mechanical and Surface Characterization of Diamond-Like Carbon Coatings onto Polymeric Substrate

Joan Martí-González, Enric Bertran[1]

[1]FEMAN Group, IN2UB, Departament de Física Aplicada i Óptica, Universitat de Barcelona. C/ Martí i Franquès 1, 08028 – Barcelona (Spain)

**ABSTRACT.** In this master thesis, diamond-like carbon DLC/Cr bilayer systems, with thickness up to 1278 nm were formed on ABS, glass and Si substrates. Substrates surface were prepared by oxygen plasma cleaning process. The chromium thin film, which acts as a buffer layer, was grown by magnetron sputtering deposition. Diamond-like carbon was deposited by pulsed-DC PECVD, with methane and hydrogen as reactants. A Plackett-Burman experimental design was carried out in order to determine the influence of technological parameters of the deposition process on the thickness, deposition rate, intrinsic stress, contact angle, roughness, friction coefficient and wear rate of the obtained coatings. The independent variables were power, chamber pressure, time of deposition, total flux of the gases, composition of the reactant gases and oxygen plasma cleaning conditions. Values of intrinsic stress between 0.13-0.78 GPa were reported. Wear resistance measurements were performed by grinding calottes with a defined geometry. Low wear rate was achieved in the range of $10^{-14}$-$10^{-15}$ $m^3/Nm$. The friction of the obtained coatings under different relative humidities, ranging 20 to 80% in a nitrogen environment, was measured using a nanotribometer with tungsten carbide ball. Results showed that friction coefficient increases with the increasing relativity humidity. Values from 0.12 to 0.24 for friction coefficient were reported. . The main objective of the DLC coating is to improve the wear resistance and tribological behavior of ABS in order to increase his durability for automotive, medical, household and textile applications, among others.

I. INTRODUCTION

Diamond-like Carbon (DLC) is an amorphous material formed by $sp^3$ and $sp^2$ carbon atoms, which can also contain hydrogen. The ratio between $sp^2$ and $sp^3$ bonds and the hydrogen content has a strong effect on properties of the DLC [1]. Depending on the bonding configuration, carbon can be found in different forms, such as graphite, diamond, polymeric or different types of amorphous carbon. **Fig. 1** shows the formed structure depending on the $sp^2$:$sp^3$ ratio and hydrogen content. In the corners there are graphite ($sp^2$), diamond ($sp^3$) and hydrogen gas. The DLC films that will be discussed in this work will be a-C:H, an amorphous mixture of hydrogen and $sp^2$:$sp^3$ carbon.

DLC coatings can have low friction, high wear resistance, high hardness, chemical inertness, smoothness, low coefficient of friction, and good optical transparency [2]. This characteristics make DLC an excellent option in order to improve surface properties of different materials and protect them.

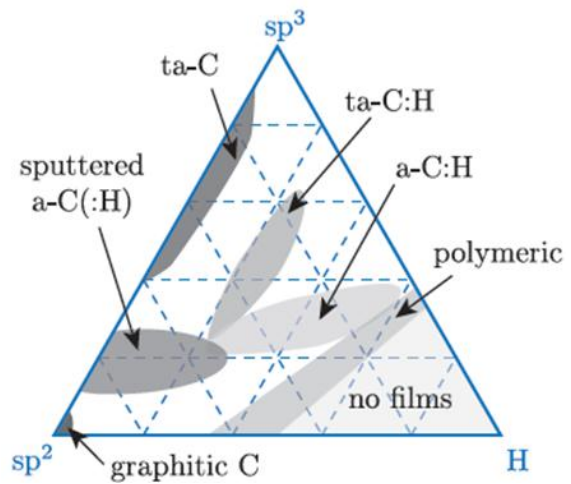

**Fig. 1.** *Ternary diagram for $sp^2$ carbon, $sp^3$ carbon and hydrogen [3].*

As has been commented, one of the main features of DLC coatings is high hardness and Young's modulus. The hardness of DLC films is in the range of 10-30 GPa, with a corresponding Young's modulus 6-10 times larger than its hardness value [4]. Compared with steel tools, hardness of DLC can be more than six time harder [5].

Another important feature of hydrogenated DLC is its tribological behaviour, which presents low wear rate and low friction coefficient. The friction coefficient comprises a wide range of values, which depends on the conditions in which it is measured (atmosphere gas, relative humidity, vacuum conditions, temperature and material of frictional surfaces), and can go from 0.001 (in dry $N_2$) up to 0.3 (in humid air) [1]. In all environments, the tribological behaviour of DLC is controlled by an interfacial transfer layer formed during friction. It consists of a friction-induced by a $sp^2$-rich layer formed on the surface of the counterpart, which is produced by the micro-debris coming from an initial abrasion stage, which shows an effective lubricating effect [6].

DLC coatings are widely used in many applications and technological fields due to its singular properties. Injection moulding, automotive and oil industries, magnetic, optical and biomedical fields are some examples. In order to accomplish this applications, it can be deposited in metal, ceramic or polymer substrates [2], [7], [8].

Techniques to deposit DLC films include physical and/or chemical processes. Some of them are suitable for the laboratory studies, while some others are preferred for the industrial production. DLC has been deposited on different substrates by many techniques. The films deposited by different methods reveal different mechanical and tribological features. Common DLC coating methods include sputtering [2], plasma enhanced chemical vapor deposition (PECVD) [5], direct ion beam deposition (direct IBD), pulsed laser deposition (PLD) and vacuum arc [4]. Choice of plasma source would depend on the intended applications. Each method is considered for a specific application and can have certain advantages and disadvantages.

One of the main drawbacks of DLC coatings is their high intrinsic compressive stress, that can achieve several GPa [4], which alters their adhesion and limits the thickness of coatings, resulting in the peeling off of the coating. Several alternatives have been studied to reduce the negative impact of internal stress. Doping DLC with metals, N or Si, post-annealing of the obtained layer or bias-graded deposition [9] are some of them. Another way consists in deposit an intermediate layer between the substrate and the DLC film in order to increase the adherence and reduce the intrinsic stress in the interface DLC/substrate. The most used intermediate layers materials are Cr, Cu, Ti, and Si [10]. Still, intrinsic stress is not the only factor that can cause the coating peeling off. Lack of surface cleaning, insufficient degassing (most polymers never stop degassing) and high temperatures of the substrate in deposition (due to the contraction when cooling) can also be factors that trigger the peeling off of the coating.

*A. Objective*

The aim of the study exposed in this report is the deposition of DLC coatings on Acrylonitrile Butadiene Styrene (ABS) substrates and the study of their mechanical properties. ABS is an amorphous thermoplastic polymer with good impact resistance, hardness and toughness, compared with other typical polymers. It is easily injecting molded and extruded, making it a good option for manufacturing a wide variety of products. The main objective of the DLC coating is to improve the wear resistance and tribological behavior of ABS in order to increase his durability for automotive, medical, household and textile applications, among others. The influence of technological parameters of DLC film deposition affecting the final properties of the coating has been evaluated. DLC films were deposited by pulsed-DC plasma-enhanced chemical vapor deposition on three different substrates (glass, silicon and acrylonitrile butadiene styrene) using an intermediate layer of Cr between substrate and DLC deposited by magnetron sputtering.

*B. Methodology*

The parameters of deposition have been determined based on previous works [3], [7], [8] in order to obtain functional layers.

A Plackett-Burman experiment design was employed to determine the effects of the typical process parameters, such as power, pressure, deposition time, gas flux, gas composition and cleaning plasma. Coating properties such as tribological behavior, intrinsic stress, wear resistance or hardness depend on both, the method and conditions of coating deposition. Therefore, the deposition process of DLC coatings represents a multi-variable system, which difficults the quantitative estimation of the effect of each parameter in the process. The relative effect of several variables on the coating have to been investigated systematically, obtaining therefore a quantitative estimation of the impact of each parameter on the process. When a huge number of process parameters are involved, the statistics-based experimental designs are used, for example in medicine or biochemistry is a widely used method [11], [12]. A usual and economical approach that provides information on the effects of single factors, but not on their interactions, is the Plackett-Burman (PB) method [13]. This method is very useful in order to determine the ruggedness of the testing, and can establish if the outcome results of an analytical procedure are affected by changes in each relevant factor and variable. PB designs employs two levels for each factor, the higher level of the range value for each parameter that we will study, expressed as "+", and the lower level, expressed as "-". For each individual experiment of the total set of experiments that a PB design involves, each factor is assigned with + or -. For example, for a set of experiments by PB design involving 6 factors (labelled A-F), one of the experiments could be: (A+ / B - / C+ / D- /E+ /F-). The rest of experiments will be combinations of this six factors with + or – values. Additionally, a center point value experiment could be done in order to improve the data processing and to collect higher quality information on the significance for each factor. The data treatment of all the results represented in Pareto charts provides the intensity of the effect of each factor for each measured property. Thickness, growth deposition, intrinsic stress, water contact angle, roughness, friction coefficient and wear rate have been characterized.

## II. EXPERIMENTAL

### A. Sample Structure

DLC/Cr coatings were deposited on three different substrates: ABS, glass and monocrystalline silicon, in order to perform different tests on each samples, as it will be explained following sections. The structure of the obtained coating is formed by two different layers, as we can see in **Fig. 2**. The first layer that had been deposited is Cr layer. The function of this intermediate layer is to improve the adhesion and reduce the intrinsic stress of the DLC layer, assuring better accommodation and less peeling off [1]. It also enhances their scientific and industrial applications for specific uses [10].

The second layer is DLC, formed by hydrogenated amorphous carbon (a-C:H). In our case, the precursor gas is methane

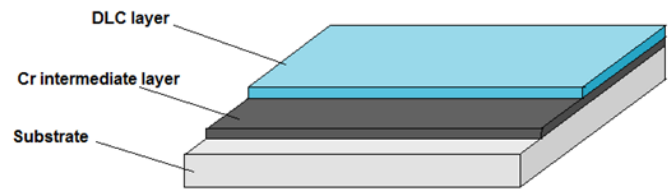

**Fig. 2.** *Scheme of the different layers of the obtained DLC/Cr/substrate coatings.*

($CH_4$). The accepted deposition mechanism of a-C:H is complicated. Several reactions take place in the plasma, which can be separated in electron-neutral, ion-neutral and neutral-neutral reactions. This reactions generates several different species, such positive ions, radicals and other hydrocarbon species. Deposition of this species to form the DLC layer involves many surface processes, such adsorption, desorption, direct incorporation of ions, reemission of H, Ion-induced incorporation of neutral radicals, adsorbed layer reactions, surface etching reactions and sputtering [14]. The most numerous species in the plasma and the ones that contribute more to the DLC growing are $CH_5^+$ and $CH_3^+$ ions [3].

### B. Preparation of the substrates

The ABS substrates were cleaned with ethanol impregnate in a cellulose tissue. Then they were cleaned with water and neutral soap in ultrasonic bath for 10 minutes. The glass substrates were cleaned with acetone in ultrasonic bath for five minutes, followed by two ultrasonic bath of isopropanol of five minutes each. The Si substrates, as they were stored in an isolated box, have been considered clean enough.

The three substrates were stuck on the cathode using Kapton® polyimide tape, special for vacuum application due to high resistivity, very low outgassing and high temperature resistance (< 269ºC). Two pen lines on top of glass substrate were made in order to measure thickness by profilometry.

### C. Deposition process

The layers were obtained by two different deposition techniques, plasma enhanced chemical vapor deposition (PECVD) and magnetron sputtering. All the deposition process was performed in the PEDRO reactor (Plasma Etching Deposition ReactOr), which is located at the FEMAN laboratory of the Faculty of Physics of the Universitat de Barcelona. In **Fig. 3** we can see a scheme of PEDRO reactor.

The reactor consists on a 50 liters main vacuum chamber, which is connected to a turbomolecular pump. An adjustable motorized conductance valve (butterfly valve) regulates the inlet cross section, between the turbomolecular pump and the main vacuum chamber, in order to keep constant the work pressure. Turbomolecular pump only works in molecular regime, where the collision between molecules is negligible. It works on the principle that gas molecules can be given a momentum in a desired direction by repeated collision with a moving solid surface. A very high spinning fan rotor hits gas molecules reaching the inlet of the pump towards the exhaust (through a rotary pump) in order to create high vacuum. The main chamber has 3 pre-vacuum chambers for cathodes (PECVD and sputtering) and sample operation, which allow the main vacuum chamber to be kept always at molecular regime (below 1 Pa). To obtain this molecular regime, rotary pumps in the pre-vacuum chambers are used. Samples are attached to the cathode in this pre-chambers, where molecular regime is reached. Then, the gate valve that isolates main vacuum chamber from pre-vacuum chamber can be opened and the samples introduced to the main vacuum chamber, in order to work under high vacuum (HV) conditions, with a mean free path greater than the dimensions of the reactor. The vacuum is measured by Pirani, Penning and capacitive pressure sensors. A radio frequency (RF) source (frequency of 13.56 MHz) (SEREN R301), with an automatic matching network, or a pulsed-DC source (ENI RPG-50 5kW Pulsed DC Plasma Generator) can be used as a power system. The cathodes have a water-cooling system in order to keep samples close to room temperature. The main chamber is connected to four gas lines and, six different gas types can be used. Each line has an automatic mass flow controller (MFC), which adjust the gas set point flow by a proportional-derivative-integration (PID) system. All the reactor and gas system is controlled and monitored with a LabView interface installed in an adjacent computer.

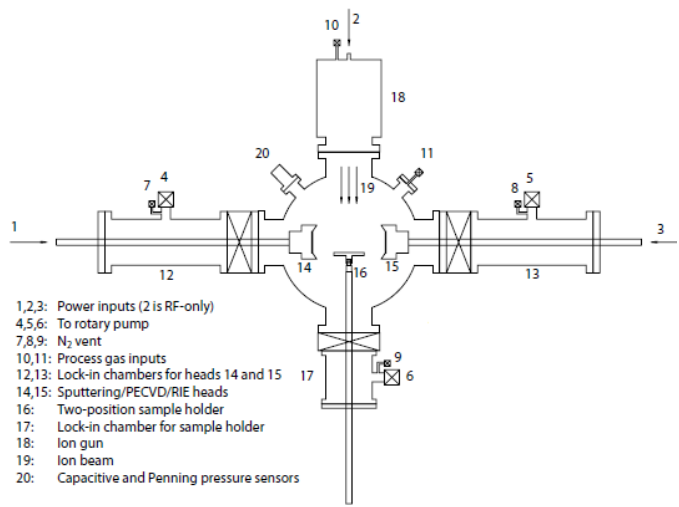

***Fig. 3.*** Schematic figure of used reactor. Used to perform cleaning plasma, magnetron sputtering and PECVD. [3].

The deposition processes involved in the production of the coatings were performed in the PEDRO reactor without breaking the vacuum inside the reactor. The sample, since the beginning to the very end of the process, was under high vacuum conditions.

*1) Cleaning plasma*

After introducing the samples in the main chamber, on the cathode, and reaching a pressure of $7 \cdot 10^{-4}$ Pa, a cleaning $O_2$ plasma was carried out by using a RF power source. Cleaning plasma has two objectives, clean the surface of small impurities and activate it. The $O_2$ plasma generates in the polymer surface functional groups such C=O, C-OH and C-O-C. These contributes to the increase of the Cr adherence to the surface, due to the enhancing of chemical reactions between Cr and polymer surfaces, resulting from the high affinity of Cr for oxygen, that forms Cr complexes and oxides that link the polymer with the Cr layer [15]. Two different oxygen plasma conditions were performed, as it was an independent variable in the Plackett-Burman experiment design. These conditions can be seen in **Table I**.

| Parameter | Plasma 1 | Plasma 2 |
|---|---|---|
| Power (W) | 50 | 20 |
| Pressure (Pa) | 20 | 10 |
| O2 flow (sccm) | 7 | 8 |
| Time (s) | 300 | 600 |

**Table I.** *Cleaning oxygen plasma conditions.*

*2) Magnetron Sputtering*

After the $O_2$ plasma cleaning, the Cr layer was deposited by magnetron sputtering technique using the RF power supply.

The magnetron sputtering technique is based on the collisions of argon ions onto a material target placed on the cathode of the reactor. After ignition of the Ar discharge, free electrons confined by the magnetron ionize the argon gas (Ar+) forming a dense Ar plasma close to the cathode. Argon positive ions bombard the target with energies around 100 eV. Argon noble gas is used because it does not to react with the target material or combine with any process gas and because produces high sputtering and deposition rates due to its molecular mass (40 amu) and relatively high kinetic moment. These positively charged argon ions are accelerated toward the negatively biased target (cathode), impacting on its surface and resulting in several Cr neutral atoms being sputtered from it. As this Cr atoms are neutrally charge, they are not affected by the electric field, and travel toward the chamber and deposit on the substrate. The cathode was located at 8 cm from our sample after the deposition rate calibration, in order to obtain a homogeneous layer on the substrate. The process conditions are summarized in **Table II**. We obtained a homogeneous layer of 223 nm of thickness and a growth rate of 0.11 nm/s.

| Parameter | Value |
|---|---|
| Power (W) | 80 |
| Pressure (Pa) | 1 |
| Ar flow (sccm) | 20 |
| Time (s) | 2000 |

**Table II.** *Cr magnetron sputtering plasma conditions.*

*3) Plasma-Enhanced Chemical Vapor Deposition*

To prepare the DLC coating, a plasma-enhanced chemical vapor deposition (PECVD) was performed. As a precursor gases, methane and hydrogen were used. Deposition conditions, which were determined by Plackett-Burman experiment design, can be seen on **Table III**. The plasma generates several species that will take part in the DLC layer growth, as explained in **section II.A**.

| Sample | Power (W) | Pressure (Pa) | CH$_4$/H$_2$ % | Flow (sccm) | Time (s) | Plasma O$_2$ |
|---|---|---|---|---|---|---|
| DLC-1 | 50 | 25 | 100 | 20 | 2400 | 2 |
| DLC-2 | 50 | 10 | 50 | 20 | 1200 | 2 |
| DLC-3 | 20 | 10 | 50 | 40 | 2400 | 2 |
| DLC-4 | 50 | 10 | 100 | 40 | 1200 | 2 |
| DLC-5 | 50 | 10 | 100 | 20 | 2400 | 1 |
| DLC-6 | 50 | 25 | 50 | 40 | 2400 | 1 |
| DLC-7 | 20 | 10 | 100 | 40 | 2400 | 1 |
| DLC-8 | 20 | 25 | 100 | 20 | 1200 | 1 |
| DLC-9 | 20 | 25 | 50 | 20 | 2400 | 2 |
| DLC-10 | 35 | 17.5 | 75 | 30 | 1800 | 1 |
| DLC-11 | 20 | 25 | 100 | 40 | 1200 | 2 |
| DLC-12 | 35 | 17.5 | 75 | 30 | 1800 | 2 |
| DLC-13 | 20 | 10 | 50 | 20 | 1200 | 1 |
| DLC-14 | 50 | 25 | 50 | 40 | 1200 | 1 |

**Table III.** *DLC deposition conditions for each sample determined by Plackett-Burman design. The frequency of pulses was kept to 100 kHz for all samples.*

The power source used for PECVD was pulsed-DC. Compared with RF source, this technique has shown an increase in deposition rate and, more important for our case, it has shown a reduction of intrinsic stress on the obtained DLC layers. This relaxation of stress is due to the alternative change of polarity of cathode bias. This characteristic of the signal enables stabilization and relaxation of the film during the positive period of each cycle ($\tau_+$), while for RF this voltage is always negative and does not allow the layer stabilization. Also, pulsed-DC has the advantage that provides more energetic ions to the cathode. For equal power and equal mean voltage, pulsed-DC shows higher peak voltages than mean voltage of RF source. The behavior of the pulsed-DC power source can be seen in **Fig. 4**. The frequency of the pulse period and the positive period ($\tau_+$) can be modified. In the experiment, constant frequency of 100 kHz (pulse period of 10 μs) and positive pulse time ($\tau_+$) of 2016 ns were used.

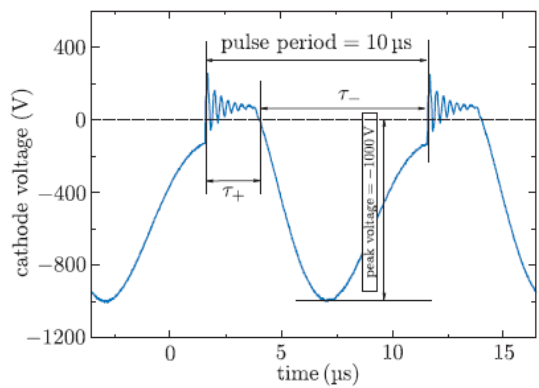

**Fig. 4.** *Pulsed-DC waveform in a 10Pa CH$_4$ discharge at 100 kHz with a positive pulse time ($\tau_+$) of 2016 ns and -1000V of peak voltage. Extracted from Corbella [14].*

### D. *Characterization techniques*

Mechanical and tribological properties of the thin film coatings were investigated using different techniques. The parameters studied were the thickness of the layer, the intrinsic stress of the film, the contact angle of water, the surface roughness, the friction coefficient and the wear resistance. All tests and analysis were performed in the FEMAN laboratories of Department of Applied Physics and Optics of the Universitat de Barcelona.

#### 1) Thickness

Thickness has been measured with confocal-interferometric microscope Sensofar PLμ 2300. Confocal microscopy is a technique that allows, in an accurate and non-destructive way, to optically section a sample in different plans, without the influence of the signal out of focus. This enhances the resolution and the contrast of the image compared to optical microscopy, and also allows the 3D reconstruction images of the observed sample. In order to measure the step height of the layers, 2 pen lines were done in the surface of the glass sample, and after DLC deposition, they were removed with ethanol, leaving a step between the substrate and the obtained DLC/Cr film. Four confocal images were taken, and the mean between all values was done in order to obtain the final result.

#### 2) Intrinsic stress

The intrinsic stress of DLC film was determined by measuring the radius of curvature of the silicon substrate before and after deposition using the confocal microscopy Sensofar PLμ 2300, and by applying the Stoney's equation [16].

$$\sigma = \frac{E_S \cdot t_S^2}{6 \cdot (1 - \nu_S) \cdot t_C}\left(\frac{1}{R} - \frac{1}{R_0}\right) \qquad (1)$$

where $\sigma$ is the intrinsic mechanical stress; $E_S$ is the Young's modulus of the substrate (c-Si); $t_S$ is the substrate thickness; $\nu_S$ is Poisson's coefficient of the substrate; $t_C$ is the film thickness; $R$ is the curvature radius of the film/substrate system, and $R_0$ is the curvature radius of the substrate without film. This is a non-destructive technique.

#### 3) Contact angle

The test was made with the optical contact angle meter KSV CAM 2000. A water droplet was dispensed onto the surface of the DLC film, and recorded with a camera. The KSV CAM 2000 software allows the calculation of the contact angle. Three droplets were recorded for glass substrate and three more for ABS substrate, obtaining the average values for each substrate. This is a non-destructive technique.

*4) Surface roughness*

Surface roughness was measured with the confocal-interferometric microscope, Sensofar PLµ 2300. Three confocal images of 1200 × 475 µm were taken for ABS substrates and three interferometric images of 2400 × 950 µm were taken for glass and Si substrates. In **Fig. 5** we can see an example. The images were treated by Sensomap software, in order to determine the roughness of each one, and the mean value was calculated for both substrates. This is a non-destructive technique.

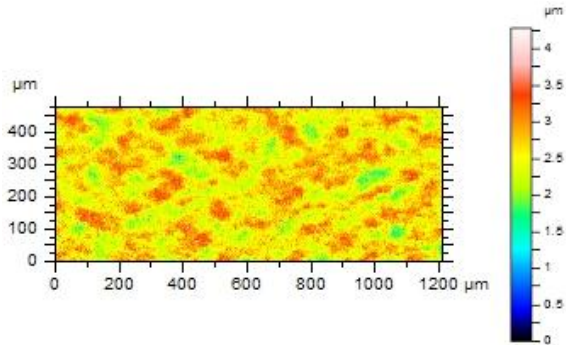

**Fig. 5.** *Confocal image of DLC-5 ABS sample. Roughness can be appreciated.*

*5) Friction coefficient*

This test was performed after the previously described ones because it is a destructive test. The friction coefficient ($\mu$) test was performed with the CSM nanotribometer, which allows different sliding tests configurations, such as pin-on-disk, ball-on-disk, ball on linear track and pin on linear track. The tests were performed with ball-on-disk method. This method consist on a tungsten carbide ball that enter in contact with the surface, with an applied normal load. The sample spins over itself and the WC ball, which is displaced from the axis of rotation, make a circular groove. The software calculates the friction coefficient using the following relation:

$$\mu = \frac{F_t}{N} \qquad (2)$$

where $F_t$ is the tangential force and $N$ the normal load. The tests were carried out with a normal load of 100 mN, 2 mm diameter WC ball, with a 1mm radius circle groove, 400 laps and a linear speed of 1 mm/s.

The tests were performed in 20, 50 and 80% of relative humidity (RH). In order to control humidity of friction tests, a self- regulated controller was used. The system allows a precise setting of RH of $N_2$ + $H_2O$ atmospheres. $N_2$ gas is bubbled in a water bath with a flow controlled by means of a mass flow controller (MFC). Temperature control of water allows different degrees of gas humidification and therefore ranges of relative humidity settings. Humid nitrogen is mixed with dry nitrogen. Finally, a RH transducer is used to regulate humid $N_2$ flow whose dynamic correction is carried out by a PID controller integrated into the MFC unit.

*6) Wear Resistance*

Wear resistance was measured using Calotest Compact unit CSM Instruments. This was a destructive test. A steel ball of 30 mm diameter spin over a sample inclined 60º. A waterbase suspension of diamond (0-0.2 µm) was uses as the grinding medium. All measures were performed at 60 rpm, 20 s, normal load of 1.07 N, 29ºC and 33% RH. The wear rate can be obtained with the following equation:

$$w = \frac{V}{s\,F} \qquad (3)$$

where $V$ is the crater volume, $s$ is the length of the path travelled by the ball, and $F$ is the normal force.

The volume of the crater was determined by confocal microscopy and the Sensomap software, which allows to obtain the exact value of the volume carved in the film.

### III. RESULTS AND DISCUSSION

#### A. Thickness and growth rate

In **Fig. 6**, the Pareto charts indicate the standardized effect of process parameters on the thickness and growth rate of the obtained DLC coatings. The length of the bars shows the degree of influence of each parameter in the obtained result and, the colour of the bar shows the positive (grey) or negative (blue) effect on the analyzed magnitude.

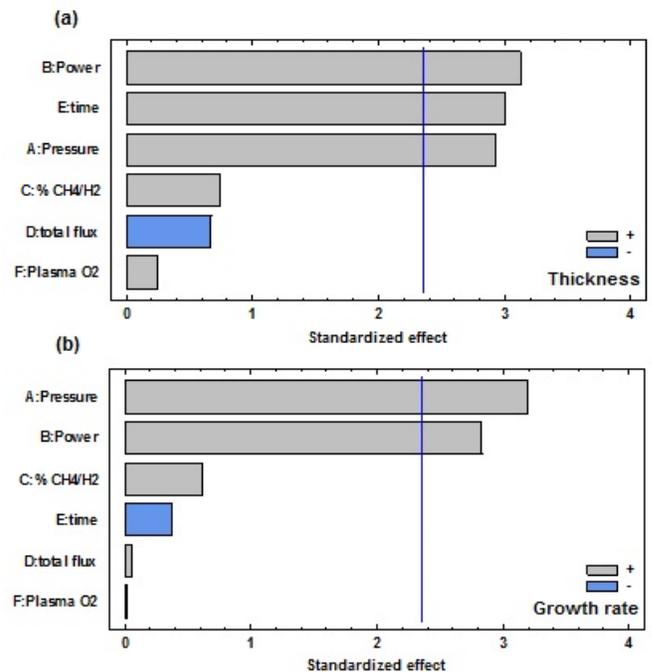

**Fig. 6.** *Pareto charts for thickness (a) and growth rate (b).*

As we can observe in Figure 6a and 6b, the Pareto charts show that an increase of power and pressure of the PECVD process has a positive, statistically significant effect on thickness and growth rate. Obviously, time also has a significant effect on thickness. As these two parameters are directly related (growth rate is thickness divided by time of deposition) it was expected that the influential parameters were the same. In contrast the other three parameters did not have a remarkable effect on thickness and growth rate. The resulting values are listed in **Table IV**.

| Sample | thickness (nm) | growth rate (nm/s) | Intrinsic stress (GPa) |
|---|---|---|---|
| DLC-1 | 1287 ± 30 | 0.54 ± 0.01 | 0.78 ± 0.02 |
| DLC-2 | 243 ± 5 | 0.20 ± 0.01 | 0.13 ± 0.02 |
| DLC-3 | 362 ± 33 | 0.15 ± 0.01 | 0.22 ± 0.09 |
| DLC-4 | 333 ± 7 | 0.28 ± 0.01 | 0.18 ± 0.02 |
| DLC-5 | 771 ± 46 | 0.32 ± 0.02 | 0.67 ± 0.06 |
| DLC-6 | 997 ± 134 | 0.42 ± 0.06 | 0.49 ± 0.13 |
| DLC-7 | 288 ± 24 | 0.12 ± 0.01 | 0.26 ± 0.08 |
| DLC-8 | 371 ± 28 | 0.31 ± 0.02 | 0.22 ± 0.08 |
| DLC-9 | 497 ± 25 | 0.21 ± 0.01 | 0.36 ± 0.05 |
| DLC-10 | 487 ± 53 | 0.27 ± 0.03 | 0.36 ± 0.11 |
| DLC-11 | 411 ± 54 | 0.34 ± 0.04 | 0.18 ± 0.13 |
| DLC-12 | 832 ± 97 | 0.46 ± 0.05 | 0.44 ± 0.12 |
| DLC-13 | 272 ± 11 | 0.23 ± 0.01 | 0.41 ± 0.04 |
| DLC-14 | 612 ± 61 | 0.51 ± 0.05 | 0.38 ± 0.10 |

**Table IV.** *Obtained values for thickness, growth rate and intrinsic stress of DLC/Cr system.*

The strong effect that power shows can be related with the fact that as higher the power, higher is the bias voltage, and ionization energy increases with bias voltage, what means a more chemically active plasma [9]. The pulsed regime produces high ion density because the increase of the ionization level, and this activates the break activity of the species present in the plasma. The plasma density increases, which is correlated with density of ions and species that involves the layer growth [3], [17]. Also, at constant power, if the pressure is increased, the bias voltage also increases, so both effects lead to an increase of thickness and growth rate [18].

**Table IV** shows the thickness and growth rate of all the samples. The highest growth values correspond to DLC-1 (25 Pa, 50W, 2400s), with thickness of 1287 nm and 0.54 nm/s of growth rate. In contrast, the lowest values are showed by DLC-7 (10 Pa, 20W, 1200s) with a thickness of 288 nm and 0.12 nm/s of growth rate. We can compare this values with other studies. Laborda [7] also deposited DLC onto polymers, and obtained deposition rates from 0.19 to 0.47 nm/s and thickness from 337 to 3100 nm pretty similar to our results except for a higher maximum thickness due to higher deposition time. Pantoja [5] reported a maximum growth rate of 0.9 nm/s and a minimum of 0.18 nm/s, working with pulsed-DC PECVD on steel substrates. Corbella [14] obtained values above 0.1-0.5 nm/s (on c-Si substrates) for pulsed DC PECVD with power values from 20 to 50 W.

*B. Intrinsic stress and Adhesion*

Intrinsic stress in the obtained DLC coatings varies in the range of 0.13-0.78 GPa, as we can see in Table 4. These values correspond to compressive stress. The compressive stress of the obtained thin films is a consequence of ion bombardment during DLC growth. Stress also depends on $sp^3/sp^2$ fraction of carbon bonds on the DLC, since the higher the ratio the higher the intrinsic stress in DLC films [9], [3]. This can be a major drawback in the production of DLC coatings, since it limits the thickness of the thin film that we can obtain [3]. The application of pulsed-DC technology has several advantages over other used technologies such as PECVD. Lower values of intrinsic stress can be obtained with pulsed-DC as a consequence of the relaxation of the layer during the positive pulse, in which there is no deposition. This phenomena reduces the stress of the deposited layer before the next pulse is received. Intrinsic stresses above 2-3 GPa are obtained by RF-PECVD while, with pulsed–DC PECVD, it is usual to have 0.1-0.8 GPa [4]. Pantoja [5] obtained DLC coatings with intrinsic stress values of 0.13-0.64 GPa by pulsed-DC PECVD and with Ti buffer layer on steel. Rubio [3] reported also values between 0.3-0.7 GPa with pulsed-DC PECVD. Corbella [14] obtained 1-1.5 GPa intrinsic stresses for pulsed-DC PECVD DLC obtained coatings without any buffer layer on c-Si. In **Fig. 7** we can observe the comparison between our obtained values (red) and Corbella's values. We can see the influence of the chromium layer (red vs black triangles), the influence of the frequency of the pulsed-DC and the influence of the power supply in the reduction of the intrinsic stress, being the lowest frequency of pulsed-DC with Cr buffer layer the conditions that showed lower values, as we wanted. Therefore, The Cr interlayer leads to considerable improvements in the adherence of stressed DLC coatings, as it acts as a buffer layer, reducing the intrinsic stress and the density of mechanical energy accumulated in the DLC [3]. The Cr interlayer obtained in all the samples has 223 nm of thickness with a growth ratio of 0.11 nm/s. Due to the resulting low intrinsic stresses, we had no peeling off of any sample, showing always a good adherence to the substrate. In previous experiments, with more extreme conditions, severe peeling off problems had been reported, so the deposition parameters were modified (lower power, short deposition time and higher Cr thickness) in order to improve adherence.

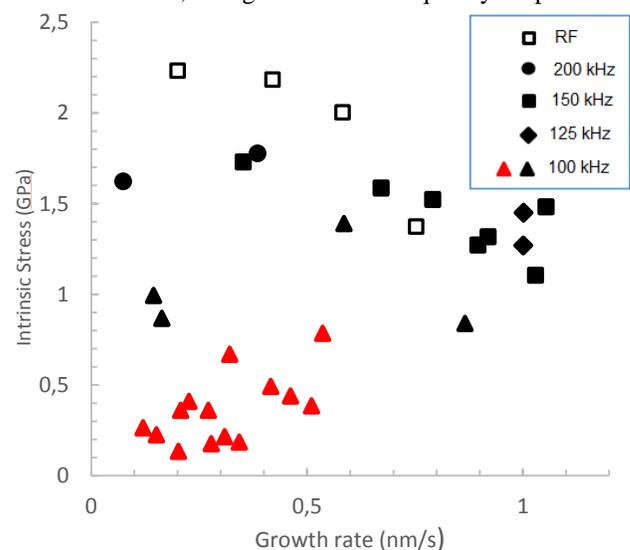

**Fig. 7.** *Intrinsic stress vs growth rate of our samples with Cr buffer layer (red) compared with Corbella [14] samples, without Cr layer. Influence of Cr buffer layer, frequency of pulsed DC and power supply can be appreciated.*

In **Fig. 8**, the Pareto chart shows that an increase of the deposition time has a positive, statistically significant effect on intrinsic stress. This means that as thicker is the layer, the higher is the intrinsic stress, as we can see in **Fig. 9 (a)** and as it has been reported in other studies [Sheeja, 19].

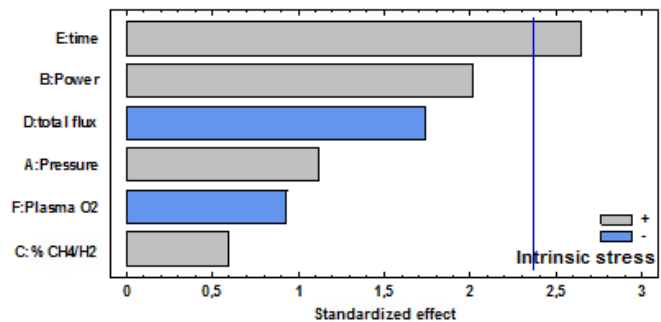

*Fig. 8. Pareto chart for intrinsic stress of the DLC/Cr depositions.*

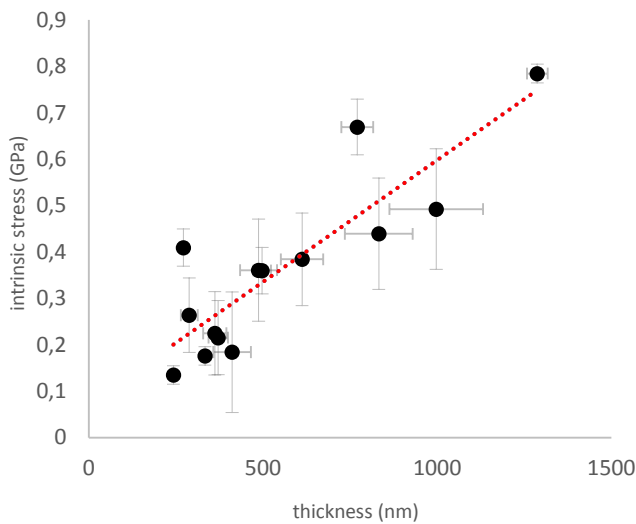

*Fig. 9. Intrinsic stress (GPa) vs thickness (nm). Intrinsic stress increases with the increasing of thickness.*

### C. Surface roughness

Roughness of 0.15-0.46 nm for glass, 0.16-0.27 nm for Si and 141-260 nm for ABS have been obtained. The Pareto chart on **Fig. 10** shows that pressure has a positive statistically significant effect on DLC roughness deposited on glass. Glass substrate roughness is average 0.171 nm and Si substrate roughness is 0.12 nm.

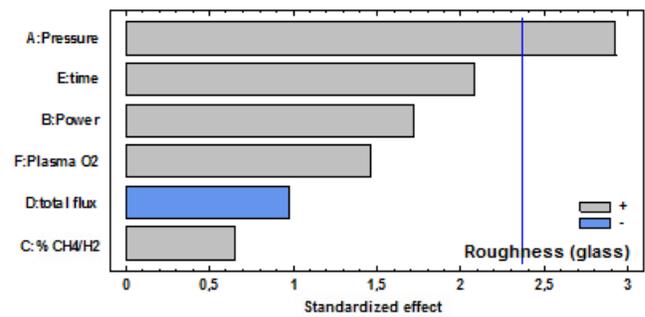

*Fig. 10. Pareto chart for roughness of DLC/Cr deposited on glass.*

As it can be seen in the glass and Si cases, surface roughness of DLC mostly depends on the surface roughness of the substrate. As DLC is amorphous, it does not present preferred growth orientations so the roughness due to the growth of the DLC layer is minimal, and can closely mimic the original surface roughness [20]. The variations in roughness induced by pressure, could be consequence of the ion bombardment increase as ion density increases with pressure. Ion bombardment can cause non-homogeneous local etching, increasing then the surface roughness of the sample and consequently the final surface roughness of the DLC coating. However, as we can see, variations on roughness are minimal, so DLC film practically mimics the surface to which is deposited to, with slightly variations. This is an important point for a good surface finish of ABS pieces or parts with a DLC protective coating.

In ABS substrates, the roughness did not show any statistically significant relation with any parameter. That can be explained by two main reasons. First, the two faces of the ABS plate, from which the samples were taken, had different average roughness (156 and 244 nm), so that leads to undermining any mathematical relationship. And finally, Roughness of ABS is so high that slightly variations in DLC growing would not significantly modify the final roughness values.

### D. Contact Angle

We can see contact angle values in **Table V**. Neither glass nor ABS substrates samples show any mathematical relation with deposition parameters. Contact angles move between 55° and 66°, so in general, no important variations are shown in contact angles of the different samples. Contact angles of glass and ABS substrates are 12.41° and 66.7°, respectively. **Fig. 11** shows the contact angle of DLC-6 ABS sample.

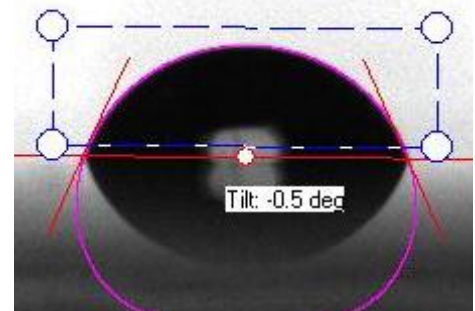

*Fig. 11. Water contact angle for DLC-6 substrate sample.*

| Sample | θ glass (°) | θ ABS (°) |
|---|---|---|
| DLC-1 | 59.2 ± 0.6 | 60.1 ± 0.8 |
| DLC-2 | 56.0 ± 1.2 | 57.6 ± 1.1 |
| DLC-3 | 62.8 ± 2.6 | 58.4 ± 4.4 |
| DLC-4 | 61.4 ± 1.9 | 62.6 ± 1.0 |
| DLC-5 | 63.4 ± 0.7 | 64.9 ± 0.5 |
| DLC-6 | 58.9 ± 2.3 | 65.7 ± 2.3 |
| DLC-7 | 58.0 ± 0.7 | 59.2 ± 2.0 |
| DLC-8 | 64.5 ± 0.9 | 61.5 ± 1.3 |
| DLC-9 | 55.1 ± 0.7 | 58.8 ± 0.6 |
| DLC-10 | 60.9 ± 0.8 | 61.8 ± 3.8 |
| DLC-11 | 61.5 ± 2.1 | 64.8 ± 4.2 |
| DLC-12 | 56.9 ± 1.2 | 59.5 ± 1.6 |
| DLC-13 | 57.4 ± 2.3 | 60.5 ± 2.9 |
| DLC-14 | 59.5 ± 4.0 | 62.4 ± 1.7 |
| Glass substrate | 12.41 ± 0.6 | |
| ABS substrate | 66.7 ± 2.4 | |

**Table V.** *Contact angle values for all the samples*

Zhang [9] Obtained contact angles from 60º to 80º using DC magnetron sputtering from graphite target to obtain DLC. In order to have superhydrophobic properties, for self-cleaning and anti-adherent surfaces, DLC can be doped by metals or fluor in order to increase the contact angle to values above 150º.

### E. Friction coefficient

The study of tribological behavior of DLC coatings was performed by three tests for each sample, at 20, 50 and 80% of relative humidity (RH), respectively. In **Fig. 12**, a typical plot obtained with the nanotribometer is shown. The oscillations that can be seen are the result of the oscillating movement of the sample, as it is spinning continually during the test process. Friction coefficient ($\mu$) in the obtained DLC coatings varies from 0.12 to 0.195 for 20% RH, from 0.14 to 0.23 for 50% RH and from 0.13 to 0.24 for 80% RH. Laborda [7] obtained, with 45% RH, friction coefficients of 0.16-0.30 on different polymer substrates with similar deposition conditions and characterization methods. Cuong [8] obtained coefficients of 0.20-0.30 for DLC coatings on polycarbonate substrate deposited by PECVD. Therefore, we obtained slightly lower values of friction coefficient, which shows that the obtained DLC coating presents a more suitable tribological behavior for the proposed applications than other similar obtained DLC coatings. In the studied case of DLC on ABS, like on other polymers, the surface hardness of substrate has an important role on the resulting friction coefficient value, because the plastic deformation, with very poor elasticity, induced by the contact requires a significant energy provided by the relative movement of sliding surfaces, which increases the work done by friction.

In our case, the deformation of the ABS surface was avoided by the use of a relatively thick Cr layer that introduces rigidity and elasticity to the system. In addition, the use of pulsed-DC plasma provides low stress DLC films. As a consequence, the use of Cr interlayer and low stress DLC provides stable and thicker films of elastic and hard DLC on ABS substrate, avoiding the ABS deformation and significantly decreasing its friction coefficient.

As it can be seen in the Pareto chart for friction at 20% and 50% RH (**Fig.13**), as power and time increases, friction coefficient decreases. This could be related to the increase of hardness and Young modulus with power [5], and the hardness effect on friction exposed above. The relation between hardness and elasticity with friction has been reported by Laborda [7].

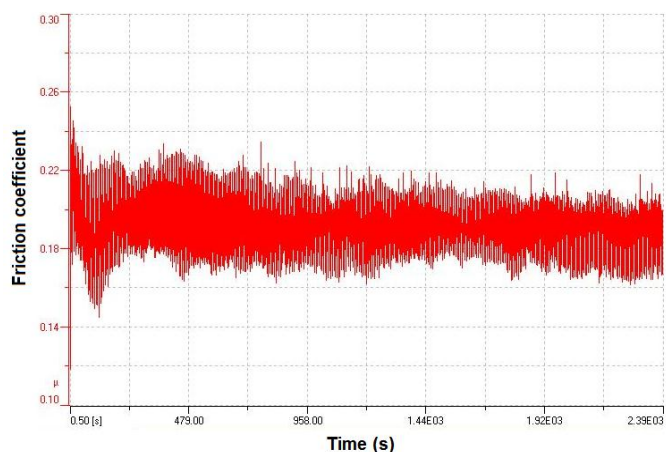

**Fig. 12.** *Plot of friction coefficient vs time for DLC-8 at 20% RH*

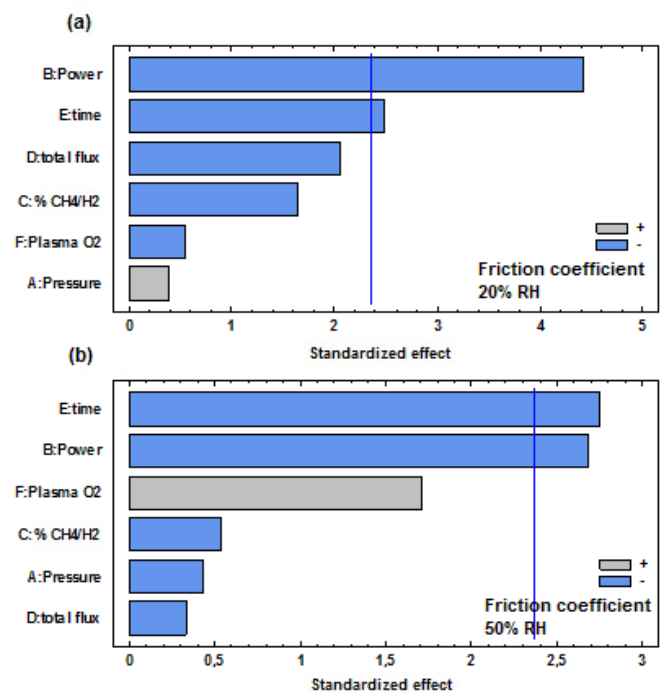

**Fig. 13.** *Pareto charts for friction coefficient at 20% RH (a) and 50% RH (b).*

n terms of relative humidity, the results presented in **Fig. 14** show a notable effect of relative humidity on the friction of DLC films. At 20% RH atmosphere, DLC samples presented the lowest friction coefficient, and it increases when % RH increases. This is caused because in a-C:H, at dry conditions, it is formed a $sp^2$-rich layer (graphitization) between the ball and DLC film.

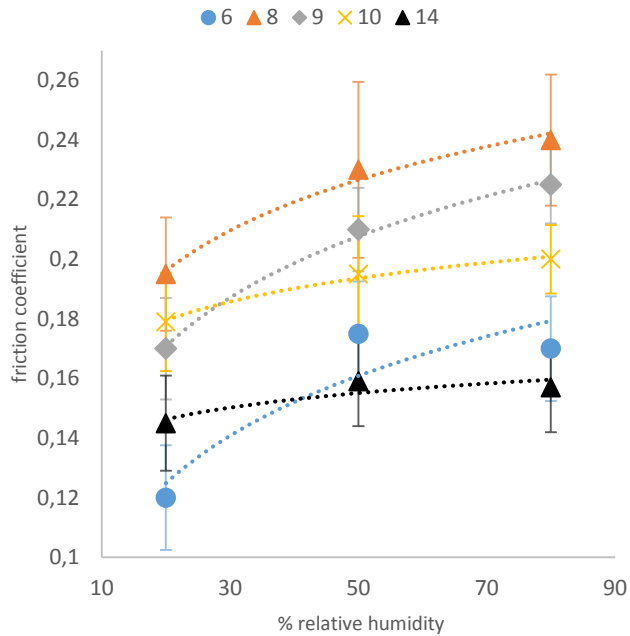

**Fig. 14.** *Friction coefficient vs % RH for samples 6, 8,9,10 and 14. The Increasing of μ with the increasing of RH can be appreciated*

This layer prevents direct contact and acts as a lubricant. As relative humidity increases, the self-lubrication capacity of the DLC decreases, due to the graphitization capacity of the DLC also decreases. This is consequence of tribo-chemical reactions in the interface that involve water molecules. Friction-induced oxidation of the DLC took place and changed the surface chemical states from C-H bonds to oxygen containing groups [20]. This effect reduces the transferred mass of carbon to the friction layer and thus reduces the lubricant effect [20]. Pantoja [5] reported similar behavior, with an increase of friction coefficient with the increasing of RH.

*F. Wear resistance*

The wear rate was evaluated from the volume of the calotte made by the steel ball using **Eq.2**. The results are shown at T**able VI**. Values from $1.22 \times 10^{-15}$ to $1.42 \times 10^{-14}$ m³/N·m are observed. This apparent huge difference is due to that some of DLC layers were worn in their totality, exposing the bottom Cr layer to suffer the wear. As Cr is softer than DLC, it suffers much more wear than DLC. In fact, the samples with higher wear rate are the ones that have thinner DLC thickness, which explains the result. This can be seen by the taken confocal 3D images, with **Fig. 15** as an example. In **Fig. 15 (a)** the debris generated during the Calotest operation can be appreciated. The obtained low wear rates are comparable to those reported by other authors obtained by similar technologies. Laborda [7] showed DLC coatings deposited on ABS with wear rates going from $5.74 \times 10^{-15}$ to $6.1 \times 10^{-16}$ m³/N·m.

| Sample | thickness(nm) | Wear rate ($10^{-15}$ m³/N·m) |
|---|---|---|
| DLC-1 | 1287 | 1.47 ± 0.15 |
| DLC-2 | 243 | 11.25 ± 1.13 |
| DLC-3 | 362 | 4.22 ± 0.42 |
| DLC-4 | 333 | 2.15 ± 0.22 |
| DLC-5 | 771 | 1.39 ± 0.14 |
| DLC-6 | 997 | 1.97 ± 0.20 |
| DLC-7 | 288 | 14.19 ± 1.42 |
| DLC-8 | 371 | 11.60 ± 1.17 |
| DLC-9 | 497 | 2.87 ± 0.29 |
| DLC-10 | 487 | 1.22 ± 0.12 |
| DLC-11 | 411 | 10.09 ± 1.10 |
| DLC-12 | 832 | 2.08 ± 0.21 |
| DLC-13 | 272 | 8.52 ± 0.85 |
| DLC-14 | 612 | 4.07 ± 0.41 |

*Table VI. Wear rate and thickness values*

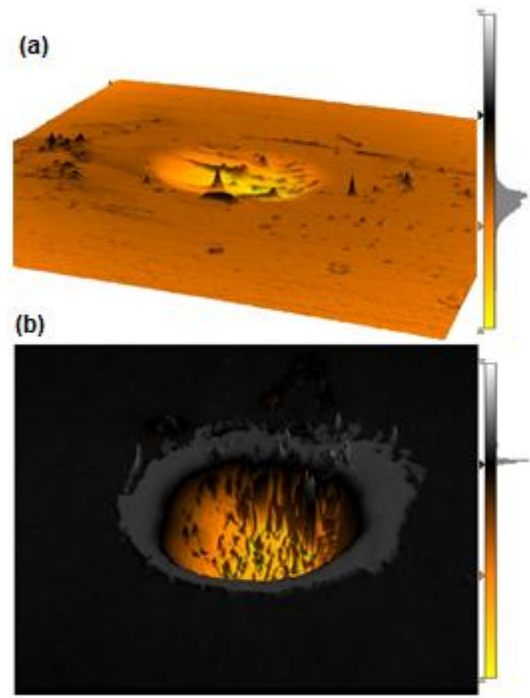

**Fig. 15.** DLC-1 sample (a) and DLC-7 sample (b) confocal 3D images. Differences between only-DLC wear (a) and DLC+Cr wear (b) can be appreciated.

Pantoja [5] reported wear rates from $10^{-13}$ to $10^{-17}$ m³/N·m. It can be seen that although our lowest wear rate is higher than the lowest value of other experiments, the values obtained are inside the range showed by similar DLC coatings.

For the possible applications of DLC coatings on ABS, these wear rate values represent an increase between four and six orders of magnitude if compared with the wear rate of untreated ABS. This circumstance, together with the other properties provided by DLC coatings, makes it advisable to apply as protective coating in a multitude of parts and plastic objects for manual use.

## IV. CONCLUSIONS

Diamond-like carbon films by pulsed-DC PECVD deposition technology with Cr buffer layer deposited by magnetron sputtering has been deposited on ABS, glass and Si substrates. Residual stress, roughness, thickness, growth rate, contact angle, friction and wear resistance were explored. In order to identify which process variables affect each investigated parameter, a Plackett-Burman experimental design was done. A set of fourteen different depositions were performed. Results showed that power, time of deposition and pressure are the main technological parameters that affect the thickness of the coatings, while pressure and power were the ones that affected the growth rate. Intrinsic stress was mostly influenced by time of deposition. It was observed that as the thickness of the layer increases, the intrinsic stress also increases. Power and deposition time were the main technological parameters affecting the friction coefficient, while roughness in glass substrates was mainly influenced by pressure. The coating properties can be optimized by modifying the technological parameters that showed a statistically significant effect on each particular property. The tribological behaviour of hyrogenated DLC shows a strong dependency with the relative humidity of the environment. Friction coefficient increased with the increase of relative humidity. The measured parameters are comparable to those reported in literature that uses similar technologies and deposition conditions. The technique showed in this work is suitable to produce hydrogenated DLC coatings on ABS substrates. As a main conclusion, low friction, high wear resistance, high adherence, smooth and stable coatings have been deposited on ABS.

## ACKNOWLEDGMENTS

The authors thanks to FEMAN group, for providing their laboratories and their useful help and suggestions during the development of project here presented. To Dr. Roger Amade, for all the support given and for helping us with the PEDRO, all the used equipment and fixing technical problems, even in his holydays. To Dr. Joan Esteve for all the help provided. And finally, a big thanks to my parents, brother, friends and my partner for all the given support through this period.